\documentclass{article}

\usepackage{arxiv}
\usepackage[utf8]{inputenc}
\usepackage[T1]{fontenc}
\usepackage{hyperref}
\usepackage{url}
\usepackage{microtype}
\usepackage{graphicx}
\usepackage{doi}
\usepackage{orcidlink}
\usepackage[style=ieee,natbib=true]{biblatex}
\usepackage{xfrac}

\usepackage{tikz}
\usetikzlibrary{mindmap}

\usepackage{authblk}
\author[1,*]{Hans Meine\orcidlink{0000-0002-7557-5007}}
\author[2]{Yongli Mou\orcidlink{0000-0002-2064-0107}}
\author[1]{Guido Prause\orcidlink{0009-0008-4273-4957}}
\author[1]{Horst Hahn\orcidlink{0000-0001-7512-5762}}
\affil[1]{Fraunhofer Institute for Digital Medicine MEVIS, Bremen, Germany}
\affil[2]{RWTH Aachen University, Germany}
\affil[*]{Correspondence: hans.meine@mevis.fraunhofer.de}

\title{On the Encapsulation of\\
Medical Imaging AI Algorithms}
\renewcommand{\headeright}{WIP -- Contributions Welcome}
\renewcommand{\undertitle}{WIP -- Contributions Welcome}

\begin{document}
\maketitle

\section*{Abstract}
In the context of collaborative AI research and development projects, it would be ideal to have self-contained encapsulated algorithms that can be easily shared between different parties, executed and validated on data at different sites, or trained in a federated manner.  In practice, all of this is possible but greatly complicated, because human supervision and expert knowledge is needed to set up the execution of algorithms based on their documentation, possibly implicit assumptions, and knowledge about the execution environment and data involved.

We derive and formulate a range of detailed requirements from the above goal and from specific use cases, focusing on medical imaging AI algorithms.
Furthermore, we refer to a number of existing APIs and implementations and review which aspects each of them addresses, which problems are still open, and which public standards and ontologies may be relevant.
Our contribution is a comprehensive collection of aspects that have not yet been addressed in their entirety by any single solution.

Working towards the formulated goals should lead to more sustainable algorithm ecosystems and relates to the FAIR principles for research data, where this paper focuses on interoperability and (re)usability of medical imaging AI algorithms.

\section{Introduction}

\subsection{Motivation}
Encapsulated algorithms should be self-contained and self-descriptive, so that they can be easily shared between parties and run quickly with minimal effort.
Algorithms that are bundled with enough metadata and that support open interfaces can be distributed via more or less public channels or "app stores".
As use cases, we consider the reuse of algorithms by colleagues or collaborators, but also running published algorithms on new datasets as a reference for benchmarking new algorithms.
A related use case are medical image computing challenges – competitions whose organizers would like to run algorithms submitted by participants on secret test data.
Finally, medical applications often come with strong data protection requirements that motivate moving algorithms to the data (often within closed hospital IT systems) instead of the other way round.
Such applications (e.g., federated or swarm learning) are another strong motivation for algorithms that do not require much manual interaction and IT knowledge to set up
but come with machine-readable meta-information that allows appropriate execution environments to run them on the desired data.

\subsection{Aspects of Encapsulation}
\begin{figure}
    \centering
    \begin{tikzpicture}[
    mindmap,grow cyclic,
    every node/.style={concept},
    concept color=orange!40,
    level 1/.append style={level distance=42mm,sibling angle=72},
    level 2/.append style={level distance=3cm,sibling angle=42},
    level 3/.append style={sibling angle=36},
]

\node{Medical\\Imaging\\AI Algorithm}
  child { node {Purpose}
    child { node {Out-of-scope use cases} }
    child { node {Intended use cases} }
    child { node {Intended users} }
    child { node {License}
      child { node { Algorithm Code } }
      child { node { Model Weights } }
      child { node { Training Code } }
    }
    child { node {Task Category} }
  }
  child { node {Output}
    child { node {I/O Interface}
      child { node {Format} }
      child { node {File\,/ Network Protocol} }
    }
    child { node {Types}
      child { node {Channel encoding} }
      child { node {Labels} }
    }
    child { node {Semantics}
      child { node {Units} }
      child { node {Description} }
      child { node {Termi\-nology Codes} }
    }
  }
  child { node {Input}
    child { node {I/O Interface}
      child { node {Format} }
      child { node {File\,/\,Network Protocol} }
    }
    child { node {Requirements}
      child { node {Image Modalities} }
      child { node {Optional Inputs} }
      }
    child { node {Parameters}
      child { node {Modes of Operation} }
    }
  }
  child { node {Source}
    child { node {Authors} }
    child { node {Institutions} }
    child { node {Code Repository} }
    child { node {Publications} }
    child { node {Training data} }
  }
  child { node {Identity}
    child { node {Name} }
    child { node {Version} }
    child { node {Date\\(last updated)} }
  }
;

\end{tikzpicture}%
    \caption{Aspects of medical imaging AI algorithms that should be modeled for proper encapsulation}
    \label{fig:aspects}
\end{figure}
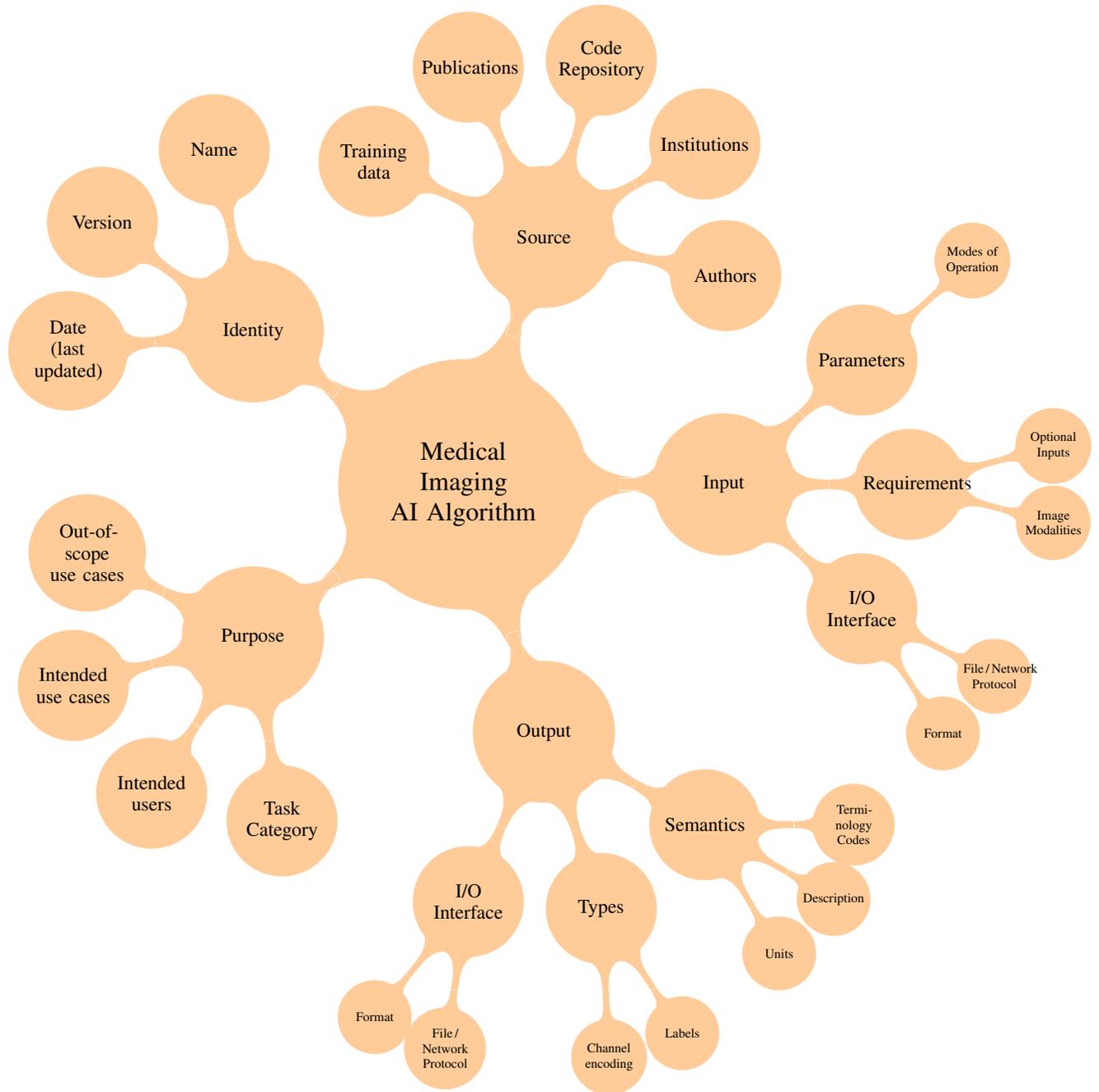
The above goals imply a set of requirements with respect to structured information that should accompany algorithms (see Figure~\ref{fig:aspects}):
\begin{itemize}
    \item Each algorithm requires an \emph{identity}. It should have a unique ID, a human-readable name, and be uniquely versioned~\cite{FAIR4RS}.
        Aspects like timestamps and checksums also fall into this category, but will not be discussed here in more detail,
        because they have been appropriately addressed in previous work and actual implementations.
    \item The \emph{source} of the algorithm should be specified, including its authors, associated institutions etc.
    \item A human-readable description of its task and purpose should be given, together with references to relevant publications that provide more details.
    \begin{itemize}
        \item In the case of data-driven algorithms (e.g., AI models), the concept of \emph{model cards}~\cite{modelcards} is continuously evolving, proposing metadata in particular on the employed datasets with a focus on transparency and ethical concerns.
        \item More recently, the FAIR4ML working group has published a metadata schema for \emph{FAIR machine learning models}~\cite{FAIR4ML}. This schema covers typical model card aspects such as ethical and social considerations, CO2 footprint, intended use, and datasets used (for training, validation, testing, and external evaluation).
    \end{itemize}
    \item The algorithm \emph{interface} needs to be described, in particular with respect to inputs and outputs, covering a number of important aspects:
    \begin{itemize}
        \item Most obviously, inputs and outputs must be specified, each with identifier and type information.
            In the case of imaging AI algorithms, typical types include \emph{images} (2D or higher dimensions),
            \emph{segmentation} results (detailed below), \emph{registration} matrices or deformation fields,
            but also input parameters (settings), possibly interaction results (click points, boxes),
            or output parameters (e.g., measurements, medical scores, estimates, or an AI model's confidence).
            Inputs may be specified as optional or mandatory.
            Types must be specified unambiguously and may include dimensionalities, bounds, ranges, or choices.
        \item In addition to the types of inputs and outputs, their technical \emph{format} must be specified.
            This is particularly relevant for (medical) images, for which a number of relevant file formats exist
            (DICOM~\cite{DICOMStandard}, NIfTI~\cite{nifti}, Analyze, NRRD, etc.)
            and it is not unusual that an algorithm supports more than one input or output format.
            Depending on the formats the data is stored or provided in and the formats supported by the algorithm
            and the execution environment, it may be possible and necessary to convert the data before or after execution.
        \item Furthermore, it may be important to describe additional \emph{semantics}, ideally also in a formalized way.
            For instance, segmentation results may be stored as label maps, but a human-readable documentation which
            integer label refers to which structure (e.g., 0=background, 1=liver tissue, 2=liver tumor tissue) does
            not allow execution environments to automatically compose algorithms, selecting, mapping, or composing labels
            as necessary to satisfy requirements.  The DICOM standard is a role model in this regard, as it provides
            unambiguous metadata linking specific encodings of geometrical structures to semantic descriptions,
            including unambiguous codes referencing external terminologies such as SNOMED CT~\cite{DICOMStandard,fedorov_dicom_2016}
            for anatomical structures, locations, or pathological findings.
        \item In order for the execution environment to be able to provide inputs to algorithms and to receive outputs from the algorithm, there must be an \emph{I/O protocol}.
            Often, this may be filesystem-based, with paths specified as program arguments or even with fixed locations in case of containerized software.
            But there are also numerous network interfaces, and algorithms may either offer their own services via such a network API or
            take the necessary information about servers to fetch the data from.
            Again, DICOM is to be mentioned as the most relevant standard for communicating medical image data,
            in particular in the form of the modern RESTful DICOMweb API (as a successor of the original, binary DIMSE protocol).
            Letting algorithms \emph{fetch} data (as opposed to passing the input data into it) is motivated by large datasets that may not be consumed as a whole,
            e.g., during machine learning applications or with histopathological whole slide scanning applications~\cite{EMPAIA}.
    \end{itemize}
    \item Algorithms may also give runtime information about the \emph{execution state}.
        This is purposely not listed among the inputs and outputs, because it follows different protocols and timings:
        Execution environments may want to display progress output (just binary "running", algorithm stages, or progress percentages) and error states and messages.
        Algorithms may decide that the input data does not match their requirements (despite prior checks by the execution environment, for instance in complex "out of distribution" situations of AI models),
        data transfers may be fail and cause execution to abort, or other runtime situations may arise and prevent the algorithm from ever producing the actual, desired output.
\end{itemize}
There are other aspects of encapsulation, in particular the term may be associated with containerization, which has become an extremely common way to facilitate algorithm exchange,
reducing the requirements on the execution environment by defining a clear interface between an algorithm bundled with all its dependencies and runtime requirements and the host.
A particularly challenging sub-aspect of this is \emph{sandboxing}, which is important when executing encapsulated algorithms from untrusted sources.

\section{Review of Existing Approaches}

\subsection{Challenge Platforms}

Online platforms that allow fair and reproducible comparisons of approaches to relevant medical tasks through so-called \emph{challenges} or competitions have at least two motivations for encapsulated algorithms:
Challenge organizers should be able to upload evaluation algorithms that systematically and automatically compute a set of performance measures for every submission.
Before the existence of such platforms, this was a tedious manual task that meant a lot of work for the organizers, even more so in case the algorithms did not run out of the box (e.g., because the submitted results were not 100\% in the required format),
and many submissions close to the deadline would result in delayed publication of the evaluation results.

Secondly, challenges nowadays often allow submission of \emph{algorithms} instead of the results from running them on test data.
This can not only reduce the workload of challenge participants as well, but it also ensures the results to be organized in the desired way, preventing many problems with the subsequent evaluation.
Most importantly, however, it allows for the test data to stay entirely hidden (including the input data), which makes the evaluation even more fair and reproducible, because participants cannot cheat or overfit to the test data as easily.


A prominent platform for challenges in medical image computing is grand-chal\-lenge.org~\cite{why_challenges}.
This open source platform allows both submissions as well as evaluation code to be uploaded as encapsulated algorithms.
Inputs and outputs are specified as part of an "interface" that declares "sockets"; each socket not only has a type, but also semantics associated with it.
For instance, tumor segmentations in CT from different timepoints would be individual sockets
(e.g., \texttt{baseline-ct-tumor-lesion-segmentaion} and \texttt{follow-up-ct-tumor-lesion-segmentation};
another example would be \texttt{pancreatic-tumor-likelihood-map}).
These sockets do not follow any previously published terminology, but are currently registered on demand via the platform support, and
at the time of writing, there are 544 such sockets defined.\footnote{\url{https://grand-challenge.org/components/interfaces/outputs/} (requires logging in)}
While grand-challenge.org is open source, the public instance is hosted on Amazon Web Services (AWS) and makes heavy use of its services,
for instance for the sandboxed execution of algorithms.

Another well-known commercial platform is Kaggle (owned by Google). While most challenges on Kaggle seem to accept submissions of solution data, there are also so-called "Code Competitions"~\cite{kaggle_competitions}.
In this case, algorithms are forced to be Jupyter notebooks which are executed on the Google Cloud Platform, and they must produce certain solution files as defined by the challenge organizers.
Code can be associated with datasets that are then provided as inputs to the notebook. Inputs and outputs are handled via the filesystem, there is no more formalized interface concept,
with the exception of an integration with Google's BigQuery service for fast access to large datasets hosted on the Google Cloud Platform.

\subsection{Federation Frameworks}

In recent years, several federated learning (or swarm learning) frameworks have been developed to enable and support a distributed approach for machine learning,
ensuring privacy and data security while still allowing for training models across different environments.
Since the main motivation of federated learning (FL) is to "bring the models to the data" instead of the other way round, this also motivates algorithm encapsulation.
Properly encapsulated, self-descriptive algorithms would require less work and less technical expertise by human operators in every participating site, allowing for faster onboarding of additional participating sites.
Ideally, also the result of a training would be immediately available as encapsulated algorithm that can be shared with colleagues or collaborators inside or outside participating institutions to be evaluated on new data.

Most FL frameworks available today do not model algorithm interfaces, but assume model and training code using the corresponding framework to be distributed to sites.



Personal Health Train (PHT) is a GO FAIR Implementation Network targeting the development of federated learning specifically for the healthcare sector.
PADME~\cite{welten_pht_2022} offers a PHT implementation heavily based on the concept of sending "trains" (docker containers) from station to station (in a network of collaborating sites), where they may analyze data or train models.
These encapsulated algorithms come with a basic interface description by means of a docker image label that describes parameters to be passed into the container as environment variables.
A basic set of types is supported, allowing to provide an interactive configuration form to the user, but no semantics or formal specification is assigned with these variables that would allow execution environments to assign values automatically.

\subsection{AI Model Repositories}

Model repositories are another motivation for encapsulating AI algorithms:
By providing rich metadata, it becomes possible to peruse large repositories, searching for algorithms that match one's available data
or that produce the required output, or that are somehow related to a task at hand.
The easier it is to execute those algorithms on one's own data, the easier it becomes to evaluate a number of candidate algorithms and to select the most appropriate one.

There are a few large model repositories, most of which are, however, not specialized for medical imaging.
Platforms like Huggingface, for instance, provide a huge number of models, but metadata is limited.
The popularity of Huggingface leads to many models being uploaded even without a human-readable description (README.md), and not every model is complete and usable as is.
It is possible to tag models with tasks, licenses, or any custom tag, but at the time of writing,
there is no "medical imaging" tag, for instance, and one cannot rely on tags being present when they would apply to the model.


The above-mentioned Kaggle platform also has a model repository built-in~\cite{kaggle_models}.
Notably, it supports variants of models (for instance, for different sizes of the same model, or different pretraining strategies) to be bundled together as a structuring mechanism.  It also comes with a predefined hierarchy of tags which allow assignment to subject areas such as "Healthcare" or "Medicine", but does not allow to describe the necessary input or the generated output in more detail than a coarse task category (e.g., "Image Classification").
A related model repository is KerasHub (part of the Keras project and also supported by Google~\cite{kerashub2024}),
but that provides even less metadata and just provides a curated subset of models from Kaggle via a Python API.

The MONAI project (Medical Open Network for AI) provides a smaller, more specialized model repository called "MONAI Model Zoo"~\cite{monai_model_zoo}.  At the time of writing, it provides 39 publicly available models.
Notably, 34 of those come as so-called "MONAI Bundles", a specification and file structure devised for distributing models with associated metadata.
This specification covers many of the aspects described above in a structured way, but is still not machine-readable; for instance, the predicted classes may be specified as "2 channels OneHot data, channel 1 is spleen, channel 0 is background".

MHub.ai (formerly known as modelhub.ai) is a model repository that tackles encapsulation of medical imaging AI in multiple ways:
Firstly, it is specific to medical and medical imaging tasks.
Secondly, it provides metadata not only on the level of the model itself, but it also describes individual inputs in a machine-readable manner.
For instance, it describes each individual output segmentation label, and
for input images, it is specified which modality (CT or MR, for instance) is expected and
whether contrast-enhanced imaging is supported or even required.
Thirdly, the aspect of optional inputs and different execution modes is solved in a structured manner, too:
The model metadata may contain individual configuration files for multiple "workflows" that define which inputs are expected and which outputs are generated in which formats~\cite{mhub_foundation_modes}.
In comparison with the model repositories mentioned above, the requirements on the model description and the technical artifacts are much higher, so it is not possible to just upload models at will, but there is a submission process that provides the necessary curation and review.
Finally, all models are wrapped in a standardized docker container (built via a continuous integration pipeline) that ensures a really consistent interface via configurable components that can perform format conversion on the fly.

MLentory~\cite{MLentory_website,MLentory_zenodo} is an interesting type of model repository:
Its stated goal is to index machine learning models from various repositories (at the time of writing including Huggingface, OpenML, and AI4Life) and to provide a unified search interface for discovering these models.
In the background, it uses an ETL pipeline to provide metadata information about the models as FDOs (FAIR Digital Objects) and makes use of the FAIR4ML schema~\cite{FAIR4ML} mentioned above.
It cannot currently provide metadata on the algorithm interface, because that's missing from the source repositories (and from the current version of the FAIR4ML schema),
but it is relevant to this work because it provides algorithm metadata in a unified schema.

\subsection{Generic Algorithm APIs}

The idea of plugin mechanisms for medical imaging algorithms is already rather old.  Even the DICOM standard originally specified a mechanism for embedding algorithms in host applications through Part 19 (PS3.19), termed "Hosted Applications"~\cite{dicom_ps3.19}.
This approach was later deprecated due to limited adoption and the complexity of maintaining a universal API across diverse clinical systems.
Modern DICOM implementations favor web services (PS3.18) and RESTful APIs for algorithm integration, which align better with contemporary software architectures.

The popular open-source medical image computing software 3D Slicer has an interesting "execution model"~\cite{slicer_execution_model} for CLI extensions (command line interface)
that allows to describe typical commandline options in a very detailed and machine-readable XML file.
This structured information is then embedded in the CLI application and used to provide either human-readable or machine-readable usage information.
The latter is used to provide GUI panels in the host application (not only 3D Slicer, but also a number of similar medical image computing platforms which have adopted this informal standard),
and the host application is then able to save input images in a compatible format, run the algorithm with the provided input data and parameters and import the output
in a way that hides the fact that the algorithm is in fact an external application.
While this approach targets interactive use of encapsulated algorithms that can be parameterized and many modern medical imaging AI algorithms can run fully automatically, it is still an interesting example for a very detailed interface description.

Many PACS vendors nowadays offer APIs to embed AI algorithms in their systems.
However, these APIs are usually not documented publicly but only available on request.


A medical image domain that comes with very specific requirements is histopathology, where whole slide imaging (WSI) produces microscopy images with six-digit extents.
Due to their huge dimensions, these images are typically stored as tiles in a resolution pyramid that allows analysis and visualization at the scale of interest (much like geomapping applications, for instance).
Nearly all approaches mentioned above rely on file system interfaces, i.e. providing input images to the algorithm directly as files in the appropriate format.
Depending on the storage location and format, this may imply temporary copying or conversion, which becomes cost-prohibitive with WSI data.
Therefore, the EMPAIA consortium has developed an open and vendor-neutral interface for AI applications in histopathology~\cite{EMPAIA}.
This app interface allows "EMPAIA Apps" (encapsulated algorithms) to request the data they need for histopathological analysis from an app server, specifying an HTTP interface to request the required image data at the required resolution and location in a tile-based manner.
The EMPAIA App Description (EAD) specifies the app interface, including input and output types such as WSI, geometrical annotations, parameters, semantic object classes, and collections.
This enables pathology workstations to present suitable apps for a given usage context, to appropriately provide input data, and to properly interpret and present their output.

\subsection{Data Management Platforms}

Another category of software are medical imaging data management and analysis platforms.  Since these also offer the integration and automatic or manual execution of algorithms, it is interesting to consider whether and how they model algorithm interfaces.

A popular open-source data management platform is XNAT~\cite{xnat}, which stores data according to an extensible data model and offers a pipeline engine that manages processing workflows triggered by data uploads or manual execution.
A "pipeline descriptor"~\cite{xnat_pipelines} describes the relation of the algorithm to the data model and an optional set of configuration parameters, but this description is from the perspective of the platform and not of the algorithm.
It should also be mentioned that a "resource descriptor"~\cite{xnat_pipelines} \emph{does} model the corresponding executable, but only its commandline arguments on a very technical level, without any semantics or other meta-information on the algorithm.
Finally, there is a "Container Service Plugin" that allows to execute docker containers in XNAT and takes a JSON command definition~\cite{xnat_json_command_definition}.
This definition covers algorithm identity and source aspects, docker execution settings, but also inputs and outputs in terms of names, types, human-readable description and routing information for the CLI.

Very similarly, the Kaapana platform~\cite{scherer_jip_kaapana_2020} offers so-called "workflows" that require specifying so-called "DAG" (based on Apache Airflow) which again describe execution pipelines from the platform perspective.
The specification is done via Python code, and the closest to an encapsulation of algorithms themselves is the definition of a DAG operator through a python class (subclassing KaapanaBaseOperator).
However, these operators also do not describe algorithm interfaces in an abstract fashion, but inputs and outputs are implicit through the code and connected through the workflow DAG.

The commercial Flywheel platform offers so-called "gears" that also allow running dockerized algorithms, either as "utility gears" (for automatic format conversions or QA purposes, for instance) that create additional data alongside their input or as "analysis gears" that also trigger a separate UI page for inspecting analysis results.  The developer documentation covering the manifest describing a gear is publicly available~\cite{flywheel_gears} and covers many aspects of encapsulated algorithms, including inputs, but excluding outputs.  Each gear may just produce an arbitrary number of output files that are automatically stored in the platform, but there does not seem to be a way to assign semantics to them.

\subsection{Relevant Standards}

While there is no single, open standard for encapsulated algorithms according to our definition yet, there are a number of relevant standards that should be mentioned in the following because they provide partial solutions or should be referenced or serve as guidance.

First and foremost, the DICOM standard is certainly the most important standard in this field.
Embracing the DICOM standard for algorithm inputs makes it possible to associate image data with the necessary metadata for their proper interpretation, but it also serves for provenance tracking and for associating algorithm outputs with the corresponding inputs, even across multiple analysis steps and with varying types of results, such as registration matrices, segmentation masks, meshes, or measurements~\cite{fedorov_dicom_2016}.

Secondly, clinical terminology systems are relevant for uniquely identifying semantic concepts that should be associated with encoded objects.
For instance, instead of just assigning a segmented structure a name, ID, or label such as "kidney", one should additionally refer to at least one coding system.  The DICOM standard does this in many places, for instance, allowing segmentation results to be linked to a code from the SNOMED Clinical Terms (where kidney would be identified as 64033007, together with the "SCT" coding scheme designator). While SNOMED offers the largest ontology, there are a number of alternatives, often more specialized, some of which are open source, such as the Foundational Model of Anatomy (FMA~\cite{fma}) or the Terminologia Anatomica~\cite{ta2}, for instance. Note that the use of SNOMED CT may require licensing in general, but several countries already pay nation-wide fees, supporting development and offering free use within the respective country. Furthermore, the DICOM standard contains a relevant subset of SNOMED terms that are free of use in the context of DICOM   

What's missing with respect to the description of algorithm inputs and outputs is a way to describe an algorithm interface –– without actually \emph{running} the algorithm, the current DICOM standard does not specify any information objects that would allow to predict, for instance, which structures \emph{would} be segmented.


With respect to a metadata schema, we have already mentioned the FAIR4ML and FAIR4RS working groups~\cite{FAIR4ML,FAIR4RS} which address aspects of machine learning models and research software,  respectively.
Image analysis algorithms are on a level in between these two, and aspects such as the inputs and outputs of algorithms, as well as their technical interfaces would require new schema concepts.
To the best of our knowledge, there is no dedicated working group or standardization effort for this specific case yet.

The concept of Research Object Crates (RO-Crates) is then a relevant standardization effort that could be used to "attach" metadata to algorithm artifacts~\cite{soilandreyes2022rocrate}.
It is not specific to algorithms, but it provides a general mechanism to bundle data (in our case the executable algorithm and any AI models contained therein) and metadata in a structured way, using JSON-LD as a serialization format.

\section{Conclusions}

While the encapsulation of algorithms has progressed a lot on the technical execution side through containerization technologies that allow bundling algorithms with all necessary dependencies, actually executing algorithms provided by others on new data requires a lot of documentation, knowledge, and human intervention in practice.  In order to facilitate algorithm exchange, discovery, automatic applicability checks, and execution, encapsulated algorithms should be fully self-descriptive in a machine-readable, standardized fashion.

We have reviewed many aspects of algorithms that should be covered by such a self-description with different use cases in mind.  We have listed existing approaches from various fields, including challenge platforms, federation frameworks, AI model repositories, and data platforms.

To summarize the state of the art, most approaches cover only basic aspects such as the identity and source of algorithms to varying degrees, usually sufficiently well, but sometimes limited in practice by lack of curation or review leading to many missing fields.
Among the approaches reviewed in this work, the mhub.ai repository for medical imaging AI models and the EMPAIA ecosystem for histopathology diagnostics offer the most detailed model of algorithm inputs and outputs,
allowing execution environments to determine automatically which algorithms can be run on which data, and to interpret and visualize results appropriately.
On the other hand, these systems require significantly more work than just uploading some trained model for packaging the algorithm to match the respective specification.  Further work on standardized algorithm interfaces and specifications for descriptions of encapsulated algorithms is needed.

\section*{Author contributions}
\textbf{HM}: Conceptualization, Investigation, Methodology, Software, Writing –– original draft, Visualization \\
\textbf{YM}: Software, Writing -- review \& editing \\
\textbf{GP}: Supervision, Project administration \\
\textbf{HH}: Conceptualization, Funding Acquisition, Supervision

\section*{Competing interests}

All authors declare that they have no competing interests.

\section*{Funding}

This manuscript was written as part of the NFDI4Health Consortium (www.nfdi4health.de). We gratefully acknowledge the financial support of the Deutsche Forschungsgemeinschaft(DFG, German Research Foundation) – project number 442326535.

This work is based on previously unpublished work on "knowledge modules" funded by the Fraunhofer Society through the AMI (Automation in Medical Imaging) and QuantMed (Quantitative Medicine) projects.

\section*{Acknowledgements}

We would like to thank Raimund Schneider and Bruno Milutin for interesting discussions on the integration of AI algorithms into PACS systems.
Furthermore, we thank Pär Kragsterman for helpful exchanges on algorithm APIs for collaboration platforms.

\printbibliography

\end{document}